# Path-memory induced quantization of classical orbits

Emmanuel Fort[a],[1] Antonin Eddi[b], Arezki Boudaoud[c], Julien Moukhtar[b], and Yves Couder[b]

[a]Institut Langevin, Ecole Supérieure de Physique et de Chimie Industrielles ParisTech and Université Paris Diderot, Centre National de la Recherche Scientifique Unité Mixte de Recherche 7587, 10 Rue Vauquelin, 75 231 Paris Cedex 05, France;

[b]Matières et Systèmes Complexes, Université Paris Diderot, Centre National de la Recherche Scientifique Unité Mixte de Recherche 7057, Bâtiment Condorcet, 10 Rue Alice Domon et Léonie Duquet, 75013 Paris, France; and

[c]Laboratoire de Physique Statistique, Ecole Normale Supérieure, 24 Rue Lhomond, 75231 Paris Cedex 05, France

[1]To whom correspondence should be addressed. E-mail: emmanuel.fort@espci.fr.

**Abstract**

A droplet bouncing on a liquid bath can self-propel due to its interaction with the waves it generates. The resulting "walker" is a dynamical association where, at a macroscopic scale, a particle (the droplet) is driven by a pilot-wave field. A specificity of this system is that the wave field itself results from the superposition of the waves generated at the points of space recently visited by the particle. It thus contains a memory of the past trajectory of the particle. Here, we investigate the response of this object to forces orthogonal to its motion. We find that the resulting closed orbits present a spontaneous quantization. This is observed only when the memory of the system is long enough for the particle to interact with the wave sources distributed along the whole orbit. An additional force then limits the possible orbits to a discrete set. The wave-sustained path memory is thus demonstrated to generate a quantization of angular momentum. Because a quantum-like uncertainty was also observed recently in these systems, the nonlocality generated by path memory opens new perspectives.

A material particle dynamically coupled to a wave packet at macroscopic scale has been discovered recently and has been shown to have intriguing quantum-like properties (1–4). The particle is a droplet bouncing on the surface of a vibrated liquid bath, and the wave is the surface wave it excites. Together they are self-propelled on the interface and form a symbiotic object. Recent investigations have shown that this "walker" exhibits a form of wave-particle duality, a unique feature in a classical system. This appears because the localized and discrete droplet has a common dynamics with the continuous and spatially extended wave. Various situations [diffraction and interference (3) and tunneling (4)], where the wave is either bounded or split, have been examined. The surprising result is that for each realization of an experiment of this type the droplet has an unpredictable individual response. However, a statistical behavior is recovered when the experiment is repeated. The truncation of the wave was thus shown to generate an uncertainty in the droplet's motion. This "uncertainty", though unrelated to Planck constant, has an analogy with the statistical behavior observed in the corresponding quantum-mechanical experiments. This characteristic has been ascribed to nonlocality. In this 2D experiment, the points of the surface disturbed by the bouncing droplet keep emitting waves. The motion of the droplet is thus driven by its interaction with a superposition of waves emitted by the points it has visited in the recent past. This phenomenon, easily observed in the wave pattern of a linearly moving walker (Fig. 1*A*), generates a path memory, a hitherto unexplored type of spatial and temporal nonlocality.

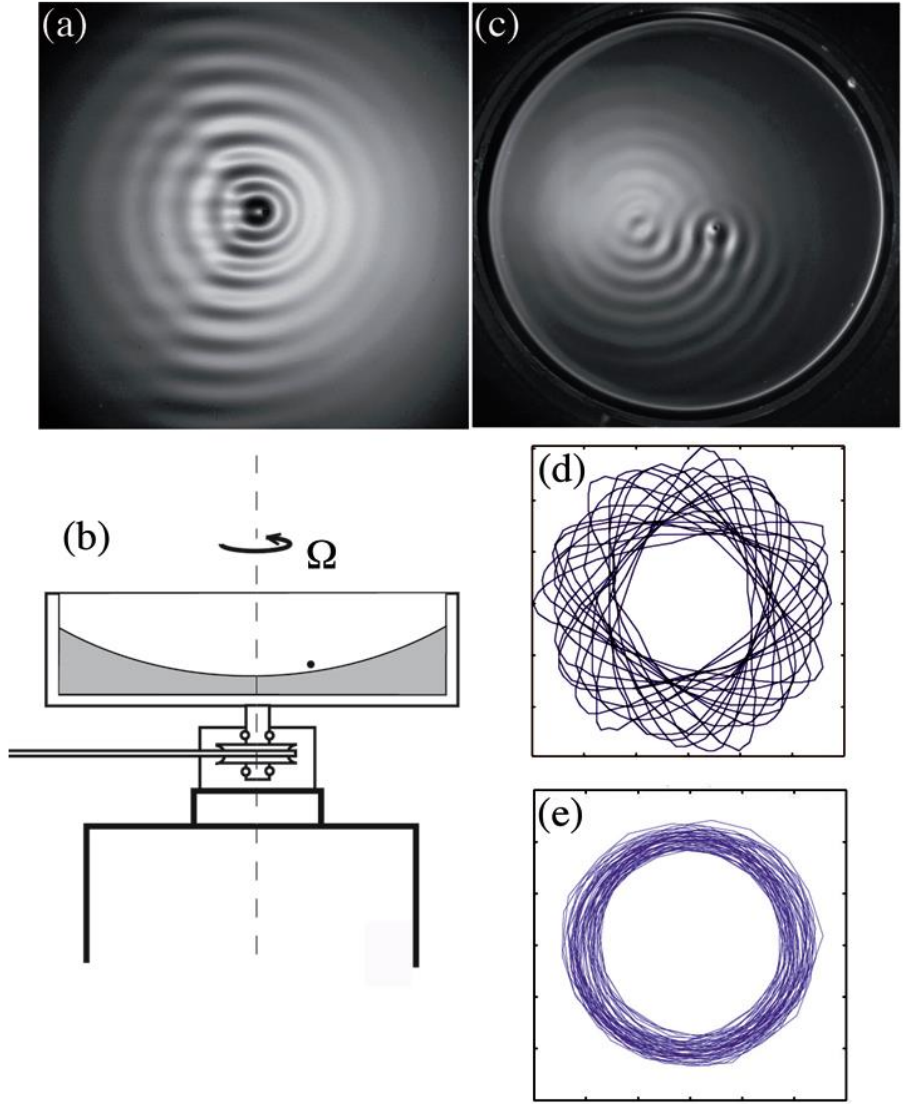

Fig. 1. *(A)* Photograph of the wave field of a walker with long-term path memory in the absence of rotation [with $\Gamma = (\gamma_m^F - \gamma_m)/\gamma_m^F = 0.018$ and $V_W = 13.7 \pm 0.1$ mm/s]. (*B*) Sketch of the experimental setup. (*C*) Photograph of the wave field of a walker of velocity $V_W = 13 \pm 0.2$ mm/s moving on a bath rotation clockwise at $\Omega = 0.9$ Rd/s. The motion is counterclockwise and the orbit $n = 3$. (*D* and *E*) The trajectory of a walker of velocity $V_W = 15.6 \pm 0.1$ mm/s on a bath rotated at $\Omega = 2.16$ Rd/s (orbit $n = 2$) as measured in the laboratory (*D*) and in the rotating frame of reference (*E*). The black segments are 2 mm long.

In the present work we show that this path-memory-driven nonlocality can lead to a form of quantization. Because the system is dissipative, its energy is imposed by a forcing at fixed frequency. Therefore there is no possible quantization of energy. For this reason we studied a possible discretization of the angular momentum of orbiting walkers similar to that of the Landau orbits of a charged particle in a magnetic field. One possibility would have been to study the trajectory of charged droplets in a magnetic field. However, for technical reasons we chose a variant that relies on the analogy first introduced by Berry et al. (5) between electromagnetic fields and surface waves.

Its starting point is the similarity of relation $\vec{B} = \vec{\nabla} \times \vec{A}$ in electromagnetism with $2\vec{\Omega} = \vec{\nabla} \times \vec{U}$ in fluid mechanics. In these relations, the vorticity $2\vec{\Omega}$ is the equivalent of the magnetic field $\vec{B}$ and the velocity $\vec{U}$ that of the vector potential $\vec{A}$. Considering a plane wave passing over a point-like vortex, Berry et al. (5) show that the phase field will exhibit defects as the wave passes on either side of a point vortex. This is the analog of the Aharonov–Bohm effect (6) where the presence of a vector potential generates a phase shift of the quantum-mechanical wave function. Although this effect is unintuitive in quantum mechanics, it receives a simple interpretation in the hydrodynamic analog where the phase shift is due to the advection of the wave by the fluid velocity. The hydrodynamic effect has been experimentally measured both in the propagation of acoustic waves (7) and in that of surface waves (8, 9) around a vortex.

In order to investigate circular orbits we use the same analogy differently. The associated forces also have similar expressions. A charge $q$ moving at velocity $\vec{V}$ in a homogeneous external magnetic field $\vec{B}$ experiences $\vec{F}_B = q(\vec{V} \times \vec{B})$. In a system rotating at an angular velocity $\Omega$, the Coriolis force on a mass $m$ moving with velocity $\vec{V}$ is $\vec{F}_\Omega = -m(\vec{V} \times 2\vec{\Omega})$, where $2\vec{\Omega}$ is the vorticity due to solid-body rotation. In classical physics, both these forces lead to orbiting motions in the planes perpendicular to $\vec{B}$ or $\vec{\Omega}$ respectively. In a magnetic field the orbit has a radius

$$\rho_L = mv/qB \qquad [1]$$

and a Larmor period $\tau_L = m/qB$. On a rotating surface a mobile particle with velocity $V$ moves on a circle of radius

$$R_C = V/2\Omega \qquad [2]$$

with a Coriolis period $T_C = 1/2\Omega$. Here we place a walker in a rotating system. If it was a normal classical particle, it would move on a circle of radius $R_C$ decreasing continuously with increasing $\Omega$. Will walkers follow this classical behavior? What is the role of the wave field induced particle path memory?

**Experimental Setup**

A sketch of our experimental system is given in Fig. 1B. The fluid is contained in a cylindrical cell of diameter 150 mm, which can be simultaneously vibrated with a vertical acceleration $\gamma = \gamma_m \sin 2\pi f_0 t$ and rotated at a constant angular velocity $\Omega$ in the range $0 < \Omega < 10$ rad/s around a central vertical axis.

In the absence of rotation, this is the setup in which bouncing droplets and walkers have already been investigated (1, 2). When a liquid bath is vertically oscillated with an increasing acceleration, there is an onset $\gamma_m^F$ beyond which the Faraday instability appears (10, 11): The entire surface of the fluid becomes unstable with formation of parametrically forced standing waves (of subharmonic frequency $f_0/2$). Our experiments are all performed below this threshold but close to it, so that the fluid surface is flat. If a small droplet (of the same fluid as the bath) is deposited on the surface, it can survive by its sustained bouncing on the interface (12). As the forcing acceleration is increased, the droplet's bouncing becomes subharmonic. It thus generates waves at the frequency of the Faraday waves. The system being tuned below, but near the threshold of this instability, these waves are almost sustained. Correspondingly, the droplet starts moving horizontally on the fluid interface. This is a symmetry-breaking phenomenon characterized by the value $\gamma_m^W$ at which the bifurcation occurs. It is worth noting that for a given fluid, walkers are observed to exist only in a finite range of droplet sizes and vibration frequencies (2). In order to investigate the wavelength dependence of the observed phenomena we used several different silicon oils of viscosities $10.10^{-3}$, $20.10^{-3}$, $50.10^{-3}$, and $100.10^{-3}$ Pa s, for which the optimum forcing frequencies are 110, 80, 50, and 38 Hz corresponding to Faraday wavelengths $\lambda_F$ = 3.74, 4.75, 6.95, and 8.98 mm, respectively. The walker's motion is recorded with a camera placed in the laboratory frame, and the trajectories in the rotating frame are reconstructed by image processing.

**Experimental Results**

**The Path Memory.** First, we address the path-memory concept. In the absence of rotation, a given droplet forms a walker in the interval $\gamma_m^W < \gamma_m < \gamma_m^F$. The bifurcation to walking being supercritical, the walker's velocity increases near the onset as $V_W \propto (\gamma_m - \gamma_m^W)^{1/2}$, then saturates at a value of the order of a tenth of $V_\phi$, the phase velocity of Faraday waves.

The global wave field results from the repeated collision of the droplet with the substrate. Each point of the surface visited by the droplet becomes the center of a localized mode of circular Faraday waves. This wave packet damps out with a typical time scale $\tau$. The transition to Faraday instability being a supercritical bifurcation, $\tau$ diverges near the threshold as $\tau \propto |\gamma_m - \gamma_m^F|^{-1}$. By tuning $\gamma_m$ we can thus control the time scale of this memory. Far from the Faraday threshold the waves are strongly damped and the wave packet has an approximately circular structure resulting from the most recent collisions of the droplet with the bath. In contrast, in the situation shown in Fig. 1A where $\gamma_m$ is close to the Faraday threshold, $\tau$ is very large and the wave field is widely extended and exhibits a complex interference structure. This wave field has an interesting relation to the Huygens–Fresnel theory of diffraction (13). The in-phase secondary sources left behind by the droplet can be considered as implementing in reality Huygens secondary sources along a wave front defined by its trajectory (14). For a rectilinear trajectory (Fig. 1A), the resulting wave pattern is thus similar to the Fresnel interference of light near the edge of a wall, the equivalent of the wall being the line of points that have not yet been visited by the walker.

We can now consider the droplet's motion. At all times it is driven by its interaction with a superposition of waves emitted at the points it visited in the past. Through the mediation of these waves, the present motion is influenced by the past trajectory. This is what we call the "wave-mediated path memory" of the system. The parameter $\tau$ measures the "time depth" of this memory.

**The Effect of Rotation.** When the experimental cell is set into rotation, the free surface of the fluid becomes parabolic (Fig. 1B). It is noteworthy that a bouncing droplet does not drift due to rotation because its bouncing on the slanted surface exactly compensates the centrifugal effect. Hence a motionless droplet remains motionless in the rotating frame. As for a walker, it walks in the rotating frame where it experiences the Coriolis force. As shown in Fig. 1C, its wave field is curled (see Movies S1 and S2). In the laboratory frame, as shown in Fig. 1D, the droplet's trajectory is epicycloidal and corresponds to the composition of an advection at an angular velocity $\Omega$ with a counterrotating Coriolis-induced orbiting. By image processing we obtain the motion in the rotating frame. It is a circular orbit of radius $R$ and angular velocity $\Omega_R^{\text{exp}}$ (Fig. 1E). In some cases the trajectory is epicycloidal in the rotating frame. In those cases we measure the amplitude of the motion having an angular velocity $\Omega_R^{\text{exp}} = 2\Omega$. When the rotation frequency is increased, the modulus of the walker velocity remains close to the value

$V_W^0$ it had without rotation, and the radius $R$ of the orbits decreases. Two distinct behaviors are observed when the system is tuned far or near the Faraday instability threshold, with short- or long-term memory, respectively.

For short-term memory, the walker moves on an orbit of decreasing radius $R_C$ (Fig. 2A) as would be expected from Eq. **2**. The measured values are, however, larger than expected and we find an excellent fit with

$$R_C^{\text{exp}} = a(V_W/2\Omega), \qquad [3]$$

where $a$ is a constant. Its value, for a given oil, is a function of the memory and ranges from 1.2 to 1.5. The origin of this factor probably results from the existence of two consecutive regimes during the bouncing. During its free flights, the droplet is submitted to a pure Coriolis effect. During the impacts, the droplet generates hollows in the interface. We believe that the interaction between the bath and the droplet during near contacts tends to oppose the bending of the trajectory. The numerical simulations presented below confirm that this hypothesis can account for the existence and value of the prefactor $a$.

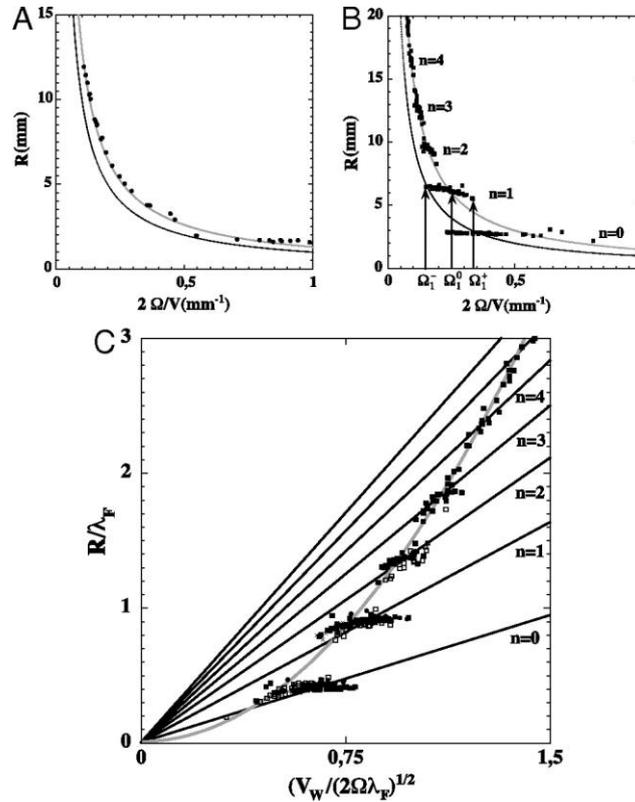

Fig. 2. (*A*) Plot of the radius $R$ as a function of $2\Omega/V_W$ in the short path-memory case. Black line: Coriolis radius $R_C = V_W/2\Omega$, gray line: fit by Eq. **3**. (*B*) Same plot for a walker in the long path-memory case (experiment performed with $\mu = 50 \cdot 10^{-3}$ Pa s, $f_0 = 50$ Hz, $\lambda_F = 6.95$ mm, and $V_W = 18$ mm/s). (*C*) Plot of the nondimensional radii $R_n/\lambda_F$ as a function of $(V_W/2\Omega\lambda_F)^{1/2}$. The results of four experiments with different oils and velocities are shown. Black squares: $\mu = 50 \cdot 10^{-3}$ Pa s, $\lambda_F = 6.95$ mm, and $V_W = 18$ mm/s. Open squares: $\mu = 20 \cdot 10^{-3}$ Pa s, $\lambda_F = 4.75$ mm, and $V_W = 12$ mm/s. Crosses: same conditions with $V_W = 8$ mm/s. Black dots: $\mu = 10 \cdot 10^{-3}$ Pa s, $\lambda_F = 3.74$ mm, and $V_W = 8$ mm/s). The gray line is the radius that would be observed with short path memory (relation **3** with $a = 1.45$). The black lines are the levels given by relation **7** with $b = 0.89$.

For long-term memory, the observed orbits are totally different (Fig. 2*B*). When Ω is increased, the radius of the orbit, instead of decreasing steadily, undergoes abrupt transitions between successive slanted plateaus. Simultaneously a slight increase of the modulus of the walker velocity $V_W(\Omega)$ is observed as the orbits become smaller.

The possible orbits being discrete, we label the plateaus by an order *n*, with *n* = 0 corresponding to the tightest orbit observed at large Ω. Each plateau crosses the continuous short path-memory curve given by Eq. **3** at a value $\Omega_n^0$ of the angular velocity (Fig. 2*B*). At these crossings, the observed orbits have the same radius $R_n^0 = R_n^{\exp}(\Omega_n^0) = R_C^{\exp}(\Omega_n^0)$ they would have had with short-term memory. A plateau extends on both sides of the curve given by relation 3 between two limiting values $\Omega_n^- < \Omega_n < \Omega_n^+$. A hysteresis is observed: The abrupt transition from one plateau to the next does not occur for the same value when Ω is increased or decreased.

On each of these plateaus the wave field exhibits a coherent global structure (see Fig. 1*C*). In this long path-memory regime, the spatial extension of the wave field exceeds the size of the orbit. Because of the circularity of the motion, the wave crests curl around and the droplet follows its own path.

We repeated the same series of experiments at four frequencies, using different oils. In each case we studied the trajectories of walkers of various velocities. The number of observed plateaus depends on the distance to onset. The closer $\gamma_m$ to $\gamma_m^F$, the larger the orbits that become discrete. Each plateau also becomes longer as hysteresis increases. Limiting ourselves to the observed plateaus we find that the radii are located on the same steps when they are rescaled by the Faraday wavelength $R_n/\lambda_F$ and plotted as a function of $(V_W/2\Omega\lambda_F)^{1/2}$ (see Fig. 2*C*).

Let us now look for an analogy between the discretization in our results and that of quantum systems. In quantum mechanics, the motion of a charge in a magnetic field is quantified in Landau levels (15). The associated Larmor radius only takes discrete values $\rho_n$ given by

$$\rho_n = \sqrt{1/\pi}\sqrt{\left(n+\frac{1}{2}\right)\frac{h}{qB}} \qquad [4]$$

with *n* = 0,1,2,…. This equation can alternatively be expressed as a function of the de Broglie wavelength $\lambda_{dB} = h/mV$. The radii then satisfy

$$\frac{\rho_n}{\lambda_{dB}} = \sqrt{1/\pi}\sqrt{\left(n+\frac{1}{2}\right)\frac{m}{qB}\frac{V}{\lambda_{dB}}}. \qquad [5]$$

The Landau orbits coincide with the classical ones when Eqs. 1 and 5 are satisfied simultaneously. Eliminating *mV/qB* gives

$$\rho_n = \frac{1}{\pi}\left(n+\frac{1}{2}\right)\lambda_{dB}, \qquad [6]$$

which corresponds to the Bohr–Sommerfeld quantization of the orbits perimeter.

We can now ask ourselves if our experimental results have some analogy with this quantization by first using the previously discussed correspondence where 1/2Ω is the analogue of the Larmor period $\tau_L = m/qB$. A bolder step is to assume that in our classical system the Faraday wavelength could play a role comparable to the de Broglie wavelength in the quantum situation. We thus try fitting our data with a dependence of the type given by

$$\frac{R_n}{\lambda_F} = b\sqrt{\left(n + \frac{1}{2}\right)\frac{1}{2\Omega}\frac{V_W}{\lambda_F}}.$$

[7]

As shown in Fig. 2*C*, an excellent fit of all the observed radii is given by this discrete set of curves. Only the dimensionless prefactor differs. Although $\sqrt{1/\pi} = 0.564$, we find the best fit for $b = 0.89$. The radii $R_n^0$ for which the classical orbits coincide with the discrete ones are those for which both Eqs. **3** and **7** are satisfied:

$$R_n^0 = \frac{b^2}{a}\left(n + \frac{1}{2}\right)\lambda_F$$

[8]

with $b = 0.89$ and $a = 1.5$ we have $b^2/a = 0.528$. This means that the diameters of the orbits are quantized:

$$D_n^0 = 2R_n^0 \simeq (n + 1/2)\lambda_F.$$

[9]

**Modeling and Numerical Simulations**

The motion of a droplet endowed with path memory on a rotating bath can be modeled and computed by numerical simulations that rely on principles previously introduced for free walkers (3). We use the simplifying approximation that the vertical and horizontal motions of the droplet are decoupled. Takeoff and landing times are thus determined only by the forcing oscillations of the liquid bath. The principle of the modeling and the notations we use are defined on Fig. 3*A*. At time $t_i$ the 2D position of the droplet is $\vec{r}_i$. The motion is iterative so that the drop will move from $\vec{r}_i$ to $\vec{r}_{i+1}$, which it will reach one Faraday period later, at time $t_{i+1} = t_i + T_F$. At each collision with the interface it undergoes a damping of its horizontal velocity due to viscous friction, but it is simultaneously given a momentum increment by its inelastic shock with the slanted oscillating surface. The direction and the intensity of the kick result from the local slope of the interface. It induces, together with the next parabolic phase, the droplet's horizontal displacement $\delta\vec{r}_i = \vec{r}_{i+1} - \vec{r}_i$.

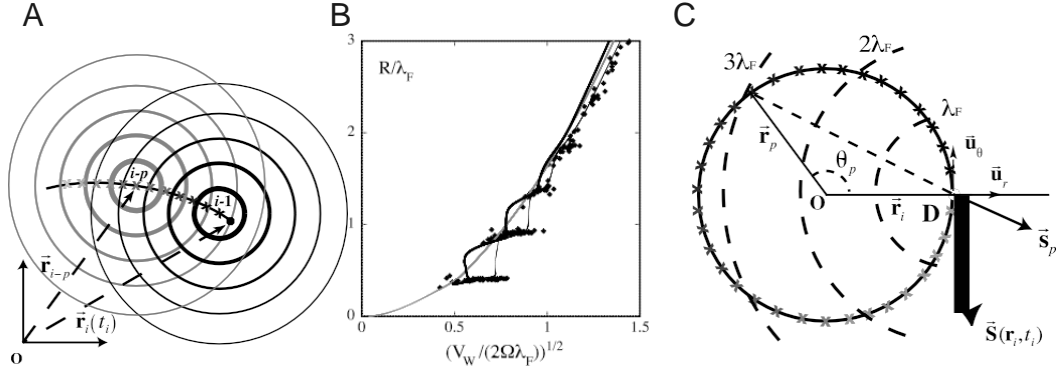

Fig. 3. (*A*) The computation of the path-memory effect. At the time $t = t_i$ the droplet, located at position $\vec{r}_i$, bounces on a surface deformed by a superposition of circular waves centered in the points where the droplet has previously bounced. (*B*) The same sketch for a circular orbit. The droplet is in *D* at $\theta = 0$ at time $t_i$ and has a clockwise motion. The temporally phase-locked wave-sources are distributed on the circle. Their temporal decay is symbolized by the fading of the gray of the crosses. Each source generates beneath the droplet a slope $\vec{s}(p)$. The norm of $\vec{s}(p)$ oscillates with the distance between the source and the droplet. The sum of these contributions is $\vec{S}(\vec{r}_i,t_i)$. In the case of a long memory, the sources are around the whole orbit. The circles centered in *D* (dashed lines) show the sources of waves reaching *D* with the same phase. (*C*) Evolution of the orbit radius obtained in the simulation with short path memory ($\tau = 10 T_F$, gray line) and long path memory ($\tau = 30 T_F$, black line), respectively. The thin and thick black lines correspond to increasing and decreasing $\Omega$, respectively; the black diamonds are the experimental data points for $\mu = 50.10^{-3}$ Pa s, $f_0 = 50$ Hz, and $V_W = 18$ mm/s.

The local surface height $h(\vec{r},t_i)$ results from the superposition of previously emitted waves as sketched in Fig. 3*A*. At each bounce in $\vec{r}_i$, a circular wave is emitted and this emission is sustained during a typical time $\tau$. The topography of the surface thus results from the superposition of these in-phase waves and contains a memory of the droplet's trajectory. In the simulations, each emitted wave is sinusoidal with the same wavelength $\lambda_F$ and phase $\phi$. The waves being 2D, their amplitudes decrease as the square root of the distance to the source. The relative surface height $h(\vec{r},t_i)$ at position $\vec{r}$ and time $t_i$ is given by

$$h(\vec{r},t_i) = \sum_{p=i-1}^{-\infty} \mathrm{Re}\, \frac{A}{|\vec{r}-\vec{r}_p|^{1/2}} \exp-\left(\frac{t_i - t_p}{\tau}\right)$$

$$\times \exp-\left(\frac{|\vec{r}-\vec{r}_p|}{\delta}\right) \exp i\left(\frac{2\pi|\vec{r}-\vec{r}_p|}{\lambda_F} + \phi\right). \quad [10]$$

where $\vec{r}_p$ is the position of a previous impact that occurred at time $t_p = t_i - (i - p)T_F$. As time passes, the amplitude of an old source decreases with a typical damping time $\tau$. The waves are also spatially damped with a typical damping distance $\delta$. Both $\tau$ and $\delta$ can be estimated from the experiments and turn out to be independent. The damping $\delta$ of traveling waves is fixed and related to the fluid viscosity. The damping time $\tau$, the path-memory parameter, is tunable, being determined by the distance to the Faraday instability threshold.

The surface slope $\vec{S}(\vec{r}_i,t_i)$ at the point of bouncing $\vec{r}_i$ at time $t_i$ is defined by the local gradient of the height in the plane of the interface. It is given by the spatial derivative of Eq. **10**: $\vec{S}(\vec{r}_i,t_i) = \vec{\nabla} h(\vec{r}_i,t_i)$. This vector is the sum of the contributions of previous sources and can be written

$$\vec{S}(\vec{r}_i,t_i) = \sum_{p=i-1}^{-\infty} \vec{s}(\vec{r}_i - \vec{r}_p, t_i - t_p). \qquad [11]$$

In the following, for simplicity we use $\vec{s}(p) = \vec{s}(\vec{r}_i - \vec{r}_p, t_i - t_p)$ and $\vec{S} = \vec{S}(\vec{r}_i,t_i)$. It had already been shown (3) that simulations using these hypotheses are able to reproduce the walking bifurcation from an initially motionless droplet. In the walking regime the simulations also produce a map of the fluid surface height that gives the topography of the wave field. Realistic fields can be obtained for all path-memory regimes. If the walker is now in a rotating frame of reference, its Coriolis acceleration has to be taken into account in the simulations. Because there is negligible influence of the Coriolis acceleration on the radial propagation of the waves, only the droplet motion has to be corrected. At each bouncing cycle, its instantaneous velocity $\vec{v}_i$ is corrected by adding an increment $\delta\vec{v}_i = -2\vec{\Omega} \times \vec{v}_i \delta t_i$. During the contact with the bath the instantaneous velocity is weak. Realistic results are obtained by considering that the Coriolis effect acts on the droplet only for the duration $\delta t_i$ of its free flights. A correcting prefactor $a$ is then observed with values similar to those measured experimentally. These simulations where the sources are distributed on a circle (Fig. 3*B*) produce steady regimes in which the droplet moves at a constant velocity on stable circular orbits. The specific phenomena observed experimentally are recovered: (*i*) With short-term path memory (e.g., $\tau = 10 T_F$), the radius of the orbits varies continuously with $\Omega$ as expected from the classical behavior. (*ii*) With long-term path memory (e.g., $\tau = 30 T_F$), the orbit radii form discrete plateaus. These plateaus cross the continuous curve associated with the short memory trajectories at discrete values $\Omega_n^0$ (Fig. 3*C*).

It is noteworthy that when submitted to rotation the modulus of the average velocity measured in the rotating frame varies only slightly as compared to the velocity of the walker in the absence of rotation. The sign of this variation depends on the memory as will be discussed below. The simulations permit the computing of surface topographies. They reproduce accurately the experimental wave field. As an example, Fig. 4 shows the comparison between the experimental and the computed wave fields for two orbits $n = 1$ and $n = 0$ with strong path memory.

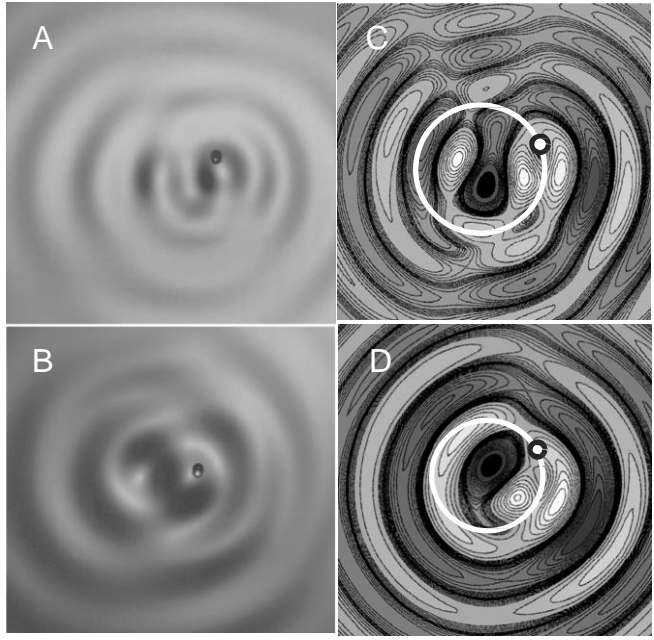

Fig. 4. Comparison of the experimental and computed central regions of the wave field. (*A* and *B*) The first two modes $n = 1$ and $n = 0$ as observed in the experiment. (*C* and *D*) The same orbits obtained in the simulation for a walker with long path memory ($\tau = 30 T_F$ and $\delta = 1.6 \lambda_F$).

**Discussion**

We now discuss the role of path memory on the dynamics of the walker. The droplet moves because it bounces on the superposition of the waves from sources on its past trajectory. All sources are synchronized by the periodic forcing. In the steady regimes where the droplet's velocity is of constant modulus the wave sources are distributed at equal intervals along the trajectory. When the walker has this rectilinear motion, all the contributions $\vec{s}(p)$ in the summation of Eq. **11** are oriented along the axis of propagation. In contrast, when the droplet has a circular motion these elementary contributions are no longer colinear as sketched in Fig 3*B*. The radial $S_r = \vec{S} \cdot \vec{u}_r$ and azimutal $S_\theta = \vec{S} \cdot \vec{u}_\theta$ components play specific roles in the dynamics of the droplet. The latter drives the propulsion of the droplet, whereas the former gives a transverse force that adds up with the Coriolis force, modifying the resulting orbit radius.

With short-term memory the active sources are located just behind the droplet on a short arc of a circle. As in the linear trajectories, these sources add up to propel the droplet. With long-time memory, the situation is more complex. The sources that have to be taken into account are distributed along the whole circle (see Fig. 3*B*). We can readily see that the contribution $\vec{s}(p)$ from past sources located immediately behind the droplet is, for purely geometrical reasons, opposite to the contribution of a source located immediately in front of the droplet. Along a closed loop of the orbit, $\vec{s}(p)$ has rotated by an angle $\pi$.

We now discuss the role of this geometric phase, which has been introduced by Berry (16), on the path-memory dynamics of the walker. We define $s_r(p)$ and $s_\theta(p)$ as the elementary contribution of a source located in $\theta_p$ to the radial and azimuthal components $S_r$ and $S_\theta$. Fig. 5 *A* and *B* show $s_\theta(p)$ and $s_r(p)$ as a function of $\theta_p$ around one orbit. They both oscillate due to the variation of the spatial phase $\phi_p = 2\pi |\vec{r}_i - \vec{r}_p| / \lambda_F$ along the trajectory. This phase increases with $\theta_p$ until $\theta_p = \pi$, then decreases symmetrically. The additional effect of the radial and azimuthal projections gives specific characteristics to $s_r(p)$ and $s_\theta(p)$. In the case of $s_\theta(p)$, the envelope is modulated by $\cos \theta_p/2$ due to the azimuthal projection. Hence, while the sources on the orbit located at $\theta_p = \pi - \theta$ have the same spatial phase $\phi_p$ as the one situated symmetrically at $\theta_p = \pi + \theta$, their contributions have opposite signs. Thus, their effect partially cancels out. The remaining contribution depends only on their relative temporal damping. This is driven by the parameter $\tau/T$, which is the relative path memory $\tau$ over one rotation period $T$.

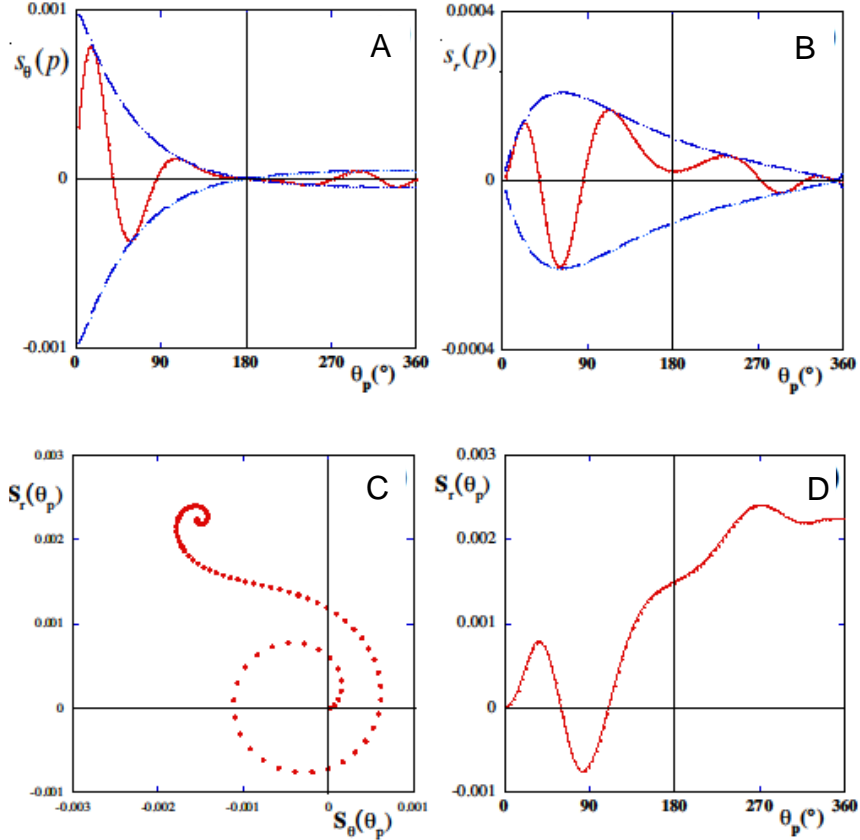

Fig. 5. (A) The individual contributions to the azimuthal slope $s_\theta(p)$ due to the sources distributed on the orbit as a function of their angular position $\theta_p$ (increasing from 0 to $2\pi$ behind the drop). (B) Same graph for the individual radial slopes $s_r(p)$. (C) The complex plane construction of a Cornu-like spiral gives the summation of the $s_r(p)$ and $s_\theta(p)$ when $\theta_p$ varies from 0 to $2\pi$ (first turn). (D) The resulting variation of the slope $S_r$ as a function of $\theta_p$.

We find experimentally that the droplet velocity $V_W$ increases slightly with $\tau/T$. However, the two contributions $s_{\pi-\theta}$ and $s_{\pi+\theta}$ cancel out more efficiently and the sum over a single loop tends toward zero. This apparent paradox is solved when integrating over all the past sources. For increasing $\tau/T$, the number of loops that must be taken into account increases and the overall integration gives resulting larger $S_\theta$ values.

We will now focus on the radial component $s_r(p)$ to understand the additional transverse force resulting from path memory that adds up with Coriolis force (Fig. 5B). The envelope is now modulated by $\sin \theta_p/2$ due to the radial projection. Hence, the contribution of the symmetrical sources with the same spatial phase $\phi_p$ located at $\theta_p = \pi - \theta$ and $\theta_p = \pi + \theta$ are additive. The sign associated with each source depends only on its spatial phase. Along a closed loop of the orbit, $\phi_p$ increases to a maximum value for $\theta_p = \pi$ and decreases symmetrically. The phasor representation helps to understand the effect of $\phi_p$. In Fig. 5C, the cumulative contribution to the radial slope $S_r(\theta_p)$ is represented as a phasor in the complex plane, in a construction similar to that of Cornu for Fresnel diffraction (17). In Fig. 5D, we show the evolution of the real part as a function of $\theta_p$ integrated up to one orbit. The structure of this phasor graph shows an S-shaped spiral structure with an inflection point at $\theta_p = \pi$. Each turn of the spiral corresponds to the spatial phase variation of $2\pi$ along an arc length of the orbit. The resulting contribution tends to cancel out. The region of the inflection point of the spiral is related to those sources diametrically opposed to the droplet on the orbit. Their contribution does not usually cancel out. It does so only for specific orbit diameters. In case of large $\tau/T$, similar contributions are generated by each turn of the droplet.

We now study the additional force as a function of the radius. As the radius increases, the S-shaped spiral rotates so that its inflection point leads to successive positive and negative jumps of $S_r$. The oscillation has the periodicity of the spatial phase of the diametrically opposed source. It thus gives $S_r$ a periodicity of $\lambda_F$ for increasing diameters. Fig. 6A shows $S_r$ as a function of the vorticity. It is characterized by a succession of sawteeth. For a discrete set of values of the radius, $S_r = 0$ and the short-term memory orbits remain unchanged. These radii correspond to the set of values of the angular velocities $\Omega = \Omega_n^0$. In all the other situations a supplementary quantization force is exerted on the droplet so that the short-term memory orbit is modified in the case of long memory. When the angular velocity is smaller than one of the $\Omega_n^0$, but close to it, the Coriolis effect should result in an orbit of larger radius. However, the slope $S_r$ is negative so that the droplet experiences an extra centripetal force that opposes the increase of the radius. Conversely, when $\Omega > \Omega_n^0$ the slope is positive and generates a centrifugal force that opposes the radius decrease. These effects are thus responsible for the formation around $\Omega_n^0$ of a slanted plateau. From this analysis, a remarkable simplification appears for the particular case of circular orbits. Because the contribution to the radial force of most sources on the circle cancel out, the remaining effect condenses to that of a single virtual source diametrically opposed to the droplet. The fundamental relation of dynamics applied in the radial direction reduces to

$$\frac{mV_W^2}{R} = -2\frac{m}{a}\Omega V_W + A\sin\left(2\pi\frac{2R}{\lambda_F} + \phi\right). \qquad [12]$$

$m$ is the mass of the droplet and $A$ and $\phi$ are, respectively, the amplitude and the phase of the wave due to the virtual droplet. $A$ follows the average characteristics of the spatial and temporal attenuation of the sources around $\theta_p = \pi$, and $\phi$ is set constant at $\pi/2$. Fig. 6B shows the computed results of this model and the fit it provides to the experimental data of Fig. 5B. The memory effect is contained in the amplitude of $A$, which reduces to zero for low memory. We thus recover in this simple model our result that quantification is related to the diameter (see Eq. **9**) and not to the perimeter as in Bohr–Sommerfeld (see Eq. **6**). We can finally note that this quantization is a priori independent of the nature of the force that sets the walkers into orbits. If it was possible to give a droplet a large electric charge and to put it in a large enough magnetic field, the same discretization of its orbits would apply.

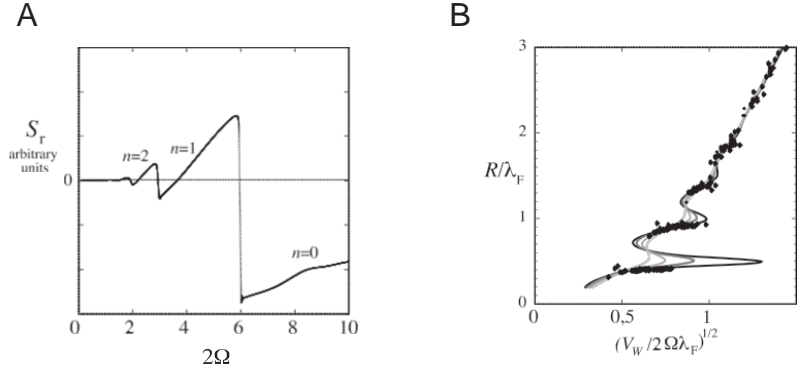

Fig. 6. (*A*) The variation of the total slope $S_r(2\pi)$ (of the sources of the last orbit) as a function of the vorticity $2\Omega$. (*B*) The fit of the experimental results by the simplified model (see Eq. **12**). Lines are darker for increasing values of the wave amplitude parameter *A* in Eq. **12**.
Previous SectionNext Section

**Conclusion**

In our experiment the droplet is driven by the wave. For this reason our system could be considered as implementing at macroscopic scale the idea of a pilot-wave considered by de Broglie (18) and Bohm (19) for elementary particles at quantum scale. However, it differs from these models by its sensitivity to history due to its wave-supported path memory. The incentive to undertake the present experiments was to investigate the constraints imposed by this memory on the possible trajectories. The walkers, when far from any disturbance, have rectilinear trajectories. When disturbed (be it by an interaction with a diffusive center or by an applied force), their trajectory becomes curved and what could be called a feed-forward effect appears. With long-term path memory, opened trajectories can become very complex leading to the probabilistic behaviors observed in diffraction (3) and tunneling (4). For closed trajectories only a discrete set of orbits is possible: The path memory is thus proven to be responsible for a self-trapping in quantized orbits reminiscent of the Bohr–Sommerfeld semiclassical quantization. It is a remarkable fact that a dynamics dominated by path memory is directly responsible for both uncertainty and quantization phenomena. We believe that the concept of path memory could be extended to Hamiltonian systems. Such an extension could bring about a previously undescribed framework for an approach to quantum nonlocality.

**Acknowledgments**

We are grateful to Mathieu Receveur and Laurent Rhea for their help in setting up this experiment and to Maurice Rossi and Eric Sultan for useful discussions. The authors acknowledge the support of the French Agence Nationale de la Recherche (ANR) under reference ANR Blanche 02 97/01.